\documentclass[12pt]{article}
\usepackage{graphicx}

\begin{document}

 \title{Beam Wandering in the Atmosphere: \\The Effect of Partial Coherence \\}
\author{G.P. Berman, $^{1}$\footnote{Corresponding
author: gpb@lanl.gov} A.A. Chumak,$^{1,2}$ and V.N. Gorshkov $^{1,2}$
\\[3mm]$^1$
 Los Alamos National Laboratory, Theoretical Division,\\ Los Alamos,
NM 87545
\\[5mm] $^2$  Institute of Physics of
the National Academy of Sciences\\ Pr. Nauki 46, Kiev-28, MSP 03028
Ukraine
\\[3mm]\rule{0cm}{1pt}}
\maketitle \markright{right_head}{LAUR-07-5341}

\begin{abstract}

The effect of a random phase screen on laser beam wander in a
turbulent atmosphere is studied theoretically. The method of photon
distribution function is used to describe the photon kinetics of
both weak and strong turbulence. By bringing together analytical and
numerical calculations, we have obtained the variance of beam
centroid deflections caused by scattering on turbulent eddies. It is
shown that an artificial distortion of the initial coherence of the
radiation can be used to decrease the wandering effect. The physical
mechanism responsible for this  reduction and applicability of our
approach are discussed.

\end{abstract}







\section{Introduction}

When a beam of light propagates through the turbulent atmosphere of
the Earth, it experiences random fluctuations in the refractive
index. Fluctuations of the refractive index are due to turbulent
eddies caused by stochastic variations of the temperature. The
characteristic scales of the atmospheric inhomogeneities range from
millimeters (the inner radius of the eddies, $l_0$) up to one
hundred meters (the outer radius of the eddies, $L_0$). Those
inhomogeneities which are large compared with the diameter of the
beam tend to deflect the beam, whereas those inhomogeneities which
are small compared with the beam diameter tend to broaden the beam
but not deflect it significantly. As a result we can observe a
broadened laser spot whose centroid randomly moves because of the
motion of individual eddies. The average beam radius is determined
by the overall scattering effect, i.e. by both the beam broadening
and the centroid wandering averaged over a sufficiently long time.

Beam wandering, as well as the scintillation index, is an important
characteristic of the radiation determining its utility for
practical applications (for example, for purposes of uninterrupted
laser tracking and pointing). Thus we will study here the
possibility of controlling this effect by means of artificially
decreasing the initial coherence of the radiation using a random
phase screen. This screen introduces random (spatial and temporal)
phase distortions into the wave front of the exiting beam.
Therefore, after passing the phase screen, the initially coherent
laser beam becomes partially coherent. Its coherence length, $l_c$,
in the direction perpendicular to the direction of propagation
becomes smaller than the diameter $D$ of the aperture. As a result
the initial angular spread of the beam, that is due to diffraction,
increases from $\lambda /D$ to $\lambda /l_c$, where $\lambda$ is
the wavelength of the radiation. (See, for example, Refs.
\cite{gbu}-\cite{dog}.) From the viewpoint of possible laser
applications, the beam broadening is a negative factor that reduces
the intensity of radiation field. At the same time, the wandering
effect can become smaller just due to the broadening.

The above comments concern the case of not too long propagation
paths when the diffraction broadening (which does really depend on
the partial coherence) dominates over broadening caused by the
atmospheric turbulence. But there is another important effect of the
phase screen on the statistical properties of the radiation
propagating in the atmosphere.  It is shown in papers
\cite{ban}-\cite{berArX} that the decrease of the initial coherence
may result in lowering the normalized variance of the intensity
(i.e. the scintillation index) even in the case of strong
turbulence. This effect takes place only for the case of a ``slow"
detector. The term ``slow detector" means the detector has an
integration time greater than the characteristic time of phase
variation introduced by the phase screen. Since the suppression of
the intensity fluctuations is of great practical importance, it is
also interesting to study the behavior of the beam wandering (which
is also expressed in terms of local fluctuations of the irradiance
intensity)  for the same experimental arrangement. Thus, in what
follows the importance of a random phase screen for the case of
strong turbulence will be elucidated.

To describe the effect of beam wandering, we will use here an
approach based on the photon distribution function \cite{ber},
\cite{berArX}.

\section{Theoretical description and calculations of the wander effect}

The position of beam centroid, ${\bf R}_w(z,t)$, is determined by the expression
\begin{equation}\label{one}
{\bf R}_w(z,t)=\frac {\int d{\bf r}_\perp {\bf r}_\perp I({\bf r},t)}
{\int d{\bf r}_\perp \langle I({\bf r},t)\rangle },
\end{equation}
where $\langle...\rangle$ means averaging over different
realizations of the refractive index inhomogeneities and source
fluctuations, ${\bf r}=\{ {\bf r}_\perp ,z\}$, ${\bf r}_\perp =\{
x,y\} $, the $z$-axis is along the initial direction of the beam
propagation; and the coordinate ${\bf r}=0$ corresponds to the
center of the exit aperture.

Following Ref. \cite{ber} we express the
intensity of photon flux $I({\bf r},t)$ in terms of the photon distribution function
$f({\bf r},{\bf q},t)$ as
\begin{equation}\label{two}
I({\bf r},t)=c\sum_{\bf q}\hbar \omega _{\bf q}f({\bf r},{\bf q},t),
\end{equation}
where $c$ is the speed of light in a vacuum, $\omega _q=cq$,
\begin{equation}\label{thr}
f({\bf r},{\bf q},t)=\frac 1V\sum_{\bf k}e^{-i{\bf kr}}b^+_{{\bf q}+
{\bf k}/2}b_{{\bf q}-{\bf k}/2},
\end{equation}
$V=L_xL_yL_z$ is the normalizing volume, and $b_{\bf q}^+$ and
$b_{\bf q}$ are the creation and annihilation operators of photons
with the wave-vector ${\bf q}$. The operator function $f({\bf
r},{\bf q},t)$ describes the photon density in $({\bf r,q})$ space
at time $t$. For a detailed description of the photon field in a
beam with radius $R_b$, it is sufficient to restrict the sum in Eq.
(\ref{thr}) with some value $k_0, |{\bf k}|<k_0$, where
$R_b^{-1}<<k_0<<\lambda ^{-1}$. In this case, the distribution
function obeys the kinetic equation (see Ref. \cite{ber})
\begin{equation}\label{fou}
\big \{ \partial_t +{\bf c_q}\partial_{\bf r}+{\bf F}({\bf r})\partial_{\bf q}\big \}
f({\bf r},{\bf q},t)=0,
\end{equation}
where ${\bf c_q}=\partial \omega _q/\partial {{\bf q}}$, ${\bf
F}({\bf r})= \omega _0\partial_{\bf r}n({\bf r})$, and $n({\bf r})$
is the fluctuating constituent of the atmospheric refractive index
($\langle n({\bf r})\rangle =0, |n({\bf r})|<<1$); $\omega_0=cq_0$
is the central frequency of laser radiation.

Using Eqs. (\ref{one}) and (\ref{two}), we can easily obtain the
variance of ${\bf R}_w(z,t)$ if one knows the correlation function
of the distribution function $\langle ff\rangle $. The analysis is
very simple for the case of weak turbulence (or short propagation
distance). In this case, it is convenient to use not directly Eq.
(\ref{one}), but a modified expression for it.  The following
consideration is in the spirit of Cook's approach  \cite{coo} who
has used the similarity between a parabolic equation describing
paraxial optical beam and the Schrodinger equation. The application
of the Ehrenfest's theorem has made it possible to develop an
approximate method to study beam wandering effect in \cite{coo}. In
contrast to Cook, we proceed not from Ehrenfest's theorem, but from
the definition (\ref{one}) and equation (\ref{fou}). A simple
relationship between the beam centroid displacement and the
refractive index fluctuations can be easily obtained within our
formalism based on the kinetic equation Eq. (\ref{fou})
\begin{equation}\label{fiv}
\bigg (\partial _z+{1\over c}\partial _t\bigg )^2{\bf R}(z,t)=\int d{\bf r}_\perp \sum _{\bf q}
\frac {\partial n({\bf r})}{\partial {\bf r}_\perp }f({\bf r},{\bf q},t)\bigg <
\int d{\bf r}_\perp \sum _{\bf q}f({\bf r},{\bf q},t)\bigg >^{-1}
\end{equation}
for the case of a stationary beam. Here and in what follows, the
paraxial approximation ($|{\bf q}-{\bf q}_0|<<q_0$ ) is assumed
throughout the beam trajectory.

In the lowest order with respect to fluctuating refractive index,
the dependence of $f$ on $n({\bf r})$ in (\ref{fiv}) has to be
neglected. Therefore, the variance of beam wandering is given by
\[\langle {\bf R}_w^2\rangle =\int _0^z\int _0^zdz_1dz_2(z-z_1)(z-z_2)\times \]
\begin{equation}\label{six}
 \int d{\bf r}_\perp d{\bf r}^\prime_\perp
\sum _{{\bf q}{\bf q}^\prime}\bigg < \frac {\partial n({\bf r})}
{\partial {\bf r}_\perp }\frac {\partial n({\bf r}^\prime)}{\partial {\bf r}_\perp ^\prime }\bigg >
\langle f({\bf r},{\bf q},t)f({\bf r}^\prime,{\bf q}^\prime ,t)\rangle
\bigg < \int d{\bf r}_\perp \sum _{\bf q}f({\bf r},{\bf q},t)\bigg >^{-2},
\end{equation}
where ${\bf r}=\{{\bf r}_\perp ,z_1\},{\bf r}^\prime=\{{\bf r}^\prime _\perp ,z_2\}$, and
$f({\bf r},{\bf q},t)$ satisfies Eq. (\ref{fou}) with ${\bf F}=0$. The distribution function at
time t
can be expressed via its value at the instant of photon exit from the source, $t_0$, as
\begin{equation}\label{sev}
f({\bf r},{\bf q},t)=f\big [{\bf r}-{\bf c}_{\bf q}(t-t_0),{\bf q},t=t_0\big ]=\frac 1V\sum_{\bf k}
e^{-i{\bf k}\big [{\bf r}-{\bf c}_{\bf q}(t-t_0)\big ]}b^+_{{\bf q}+{\bf k}/2}b_{{\bf q}-{\bf k}/2}|_{t=t_0},
\end{equation}
where $t-t_0=z/c$.  In what follows we will put $t_0=0$ for
simplicity.

Two independent averagings should be undertaken in the right-hand
part of Eq. (\ref{six}). These are averaging over the turbulence
configurations, $\langle nn\rangle$, and averaging over the
fluctuations of the source, including fluctuations introduced by the
random phase screen, $\langle ff\rangle $. The first of these is
determined by the known \cite{fan} expression
\begin{equation}\label{eig}
\langle n({\bf r})n({\bf r}^\prime)\rangle =\int d{\bf g}e^{-i{\bf g}({\bf r}-{\bf r}^\prime )
}\psi ({\bf g}),
\end{equation}
where the explicit term for $\psi $ is given by
\begin{equation}\label{nin}
\psi ({\bf g})=0.033C_n^2\frac {\exp[-(gl_0/2\pi
)^2]}{[g^2+L_0^{-2}]^{11/6}}.
\end{equation}
Eq. (\ref{nin}) is referred to as the von Karman spectrum.

The other averaging accounting for the effect of a ``slow" detector
is given by \cite{ber}:
\[\langle f({\bf r},{\bf q},t)f({\bf r}^\prime,{\bf q}^\prime ,t)\rangle=
\bigg (\frac {2\pi r_1^2}{VL_xL_y}
\bigg )^2\delta _{q_z,q_0}\delta _{q^\prime_z,q_0}\langle b^+b^+bb\rangle \times \]
\begin{equation}\label{ten}
\sum _{{\bf k}_\perp ,{\bf k}_\perp ^\prime }e^{-i{\bf k}_\perp \big ({\bf r}_\perp -{\bf q}_\perp z_1/q_0\big )
-i{\bf k}^\prime _\perp \big ({\bf r}^\prime_\perp -{\bf q}^\prime_\perp z_2/q_0\big )}
e^{-\big (k^2_\perp +k^{\prime ^2}_\perp \big )r_0^2/8-\big (q^2_\perp +q^{\prime ^2}_\perp \big )r_1^2/2},
\end{equation}
where $b^+,b$ are the operators of the generated mode. The effect of
the phase screen is represented in Eq. (\ref{ten}) by the parameter
$r_1$, determined via the correlation length, $\lambda_c $, of phase
variation due to the phase screen as
\begin{equation}\label{ele}
r_1^2=\frac {r_0^2}{1+2r_0^2\lambda _c^{-2}}.
\end{equation}
In the absence of a phase screen, we may set formally $\lambda
_c=\infty $. Then it follows from Eq. (\ref{ele}) that in this case
$r_1=r_0$. For any finite value of  $\lambda _c$, $r_1<r_0$ and, as
follows from Eq. (\ref{ten}), the characteristic values of the
transverse momenta of photons $q_\perp , q^\prime _\perp$ are
increased. This means that the beam becomes more divergent after
passing a phase screen .

With the known explicit terms (\ref{nin}) and (\ref{ten}), the
calculation of $\langle{\bf R}_w^2\rangle $ reduces to many-fold
integrations which can be performed straightforwardly. The result is
given by
\begin{equation}\label{twe}
\langle {\bf R}_w^2\rangle =0.066\pi ^2\Gamma \bigg ({\frac 16}\bigg )C_n^2z^{8/3}(q_0r_1)^{1/3}I_1.
\end{equation}
The dimensionless quantity $I_1$ is determined by the integral
\begin{equation}\label{thi}
I_1=\int_0^1dx(x-1)^2\big [x^2+(r_0^2/4+l_0^{\prime 2})q_0^2r_1^2z^{-2}\big ]^{-1/6},
\end{equation}
where $l_o^\prime =l_0/(2\pi )$. It  can be calculated numerically. In the limiting cases
of short and long propagation distances it is given by the following analytical
expressions:
\[I_1\approx {1\over 3}\bigg ({z\over q_0r_1}\bigg )^{1/3}\big (l_0^{\prime 2}+
{r_0^2/4}\big )^{-1/6},~ when ~~ q_0^2r_1^2z^{-2}\big (l_0^{\prime 2}+{r_0^2/4}\big )>>1;\]
\begin{equation}\label{four}
I_1\approx {27\over 40},~when ~~ q_0^2r_1^2z^{-2}\big (l_0^{\prime 2}+{r_0^2/4}\big)<<1.
\end{equation}
Usually $(r_0^2/4)>>l_0^{\prime 2}$. Then, the above criteria mean
that the diffraction broadening is smaller (upper case) and greater
(lower case) than the initial beam radius. The upper case in Eqs.
(\ref{four}) results in $\langle {\bf R}^2_w\rangle
=1.919C_n^2z^3(2r_0)^{-1/3}$, that coincides exactly with the
classic formula presented in the reference \cite{fan80}. (See Eq.
(45) there.) As we see, there is no dependence of beam wandering on
the phase screen when the propagation distance is very short. The
result is evident for this limiting case in view of the fact that
both the diffraction broadening and the broadening due to the
atmosphere turbulence is much smaller than the initial radius of the
beam. With the increase of propagation distance, $z$, or decrease of
the initial coherence, the upper case in Eqs. (\ref{four}) may
transform to the lower case which corresponds to dominating
diffraction broadening of the beam. Then the dependence $\langle
{\bf R}^2_w\rangle \sim r_1^{1/3}$ will arise. As we see $\langle
{\bf R}^2_w\rangle $ decreases with  decreasing initial coherence.
In this case the variance of the wander distance can be controlled
by a suitable choice of the random phase screen.

The situation is much more complex when the turbulence is strong.
The averaging is no longer decoupled in the manner shown in Eq.
(\ref{six}). An essential dependence of the distribution function on
turbulence takes place here. Therefore the approach based on
employing Eq. (\ref{fiv}) is not advantageous. The simplest way for
further analysis is to proceed from the initial definition of the
wandering given by Eq. (\ref{one}). The expression for the
distribution function $f({\bf r},{\bf q},t)$ is given by \cite{ber}
\begin{equation}\label{fift}
f({\bf r},{\bf q},t)={1\over V}\sum_{\bf k}e^{-i{\bf k}_\perp \big \{ {\bf r}-{\bf c_q}t+
\frac c{q_0}\int_0^tdt^\prime t^\prime {\bf F}_\perp  \big [{\bf r}(t^\prime )\big ] \big \} }
b^+_{{\bf Q}+{\bf k}_\perp /2,q_0}b_{{\bf Q}-{\bf k}_\perp /2,q_0/2}|_{t=0},
\end{equation}
where ${\bf q}=\{ {\bf q}_\perp ,q_0\} , {\bf Q}={\bf q}_\perp -\int_0^t
dt^\prime {\bf F}_\perp [{\bf r}(t^\prime)]$, and ${\bf r}(t^\prime)$ is the
trajectory of particle, which has the velocity ${\bf c}_{{\bf q}(t^\prime)}$
and is affected by the force
${\bf F}$. The initial conditions are given by
${\bf r}(t^\prime =t)={\bf r}$ and ${\bf q}(t^\prime =t)={\bf q}$.

Substituting Eq. (\ref{fift}) into the general expression
\[\int \int d{\bf r}_\perp
d{\bf r}^\prime _\perp {\bf r}_\perp {\bf r}^\prime _\perp \langle I({\bf r},t)I({\bf r}^\prime,t)\rangle ,\]
which determines the mean-square variation of the wander distance, and averaging
over phase variations introduced by the random phase screen, we arrive at
\[ \bigg (\frac {2\pi r_1^2c\hbar \omega}{VL_xL_y}\bigg )^2\langle b^+b^+bb\rangle
\sum_{{\bf k}_\perp ,{\bf q}_\perp }\sum_{{\bf k}_\perp ^\prime ,{\bf q}_\perp ^\prime}
\int d{\bf r}_\perp d{\bf r}^\prime _\perp
{\bf r}_\perp {\bf r}^\prime _\perp e^{-i{\bf k}_\perp ({\bf r} -{\bf c_q}t)
-i{\bf k}^\prime _\perp ({\bf r}^\prime -{\bf c}_{{\bf q}^\prime }t)}\]
\begin{equation}\label{sixt}
\times e^{-(k^2_\perp +k^{\prime ^2}_\perp )r_0^2/8}\bigg <e^{-(Q^2 +Q^{\prime ^2})r_1^2/2-
(ic/q_0)\int _0^tdt^\prime t^\prime \big [{\bf k}_\perp {\bf F}({\bf r}(t^\prime ))
+{\bf k}^\prime _\perp {\bf F}({\bf r}^\prime (t^\prime ))\big ]}\bigg >,
\end{equation}
where the last averaging is over random values of the refraction index. It is worth mentioning
that momenta ${\bf Q}$ and ${\bf Q}^\prime $ depend linearly on ${\bf F}$.
Therefore it is convenient to rewrite the quantity in the last angle bracket in a more convenient
equivalent form as
\[\bigg <...\bigg >={1\over (2\pi r_1^2)^2}\int \int d{\bf p}d{\bf p}^\prime
e^{i({\bf pq}+{\bf p^\prime q^\prime})-(p^2 +p^{\prime ^2})/(2r_1^2)}\]
\begin{equation}\label{seve}
\times \bigg < e^{-i\int _0^tdt^\prime \big [\big (\frac c{q_0}{\bf k}_\perp t^\prime +{\bf p}\big )
{\bf F}[{\bf r}(t^\prime )]+\big (\frac c{q_0}{\bf k}^\prime _\perp t^\prime +{\bf p}^\prime \big )
{\bf F}[{\bf r}^\prime (t^\prime )]\big ]}\bigg >,
\end{equation}
where ${\bf p}$ and ${\bf p}^\prime$ are vectors perpendicular to
the $z$-axis.

Thus, the problem is reduced to the calculation of manyiple
integral. There is a $13$-fold integration in Eq. (\ref{sixt}).
After substituting Eq. (\ref{seve}) into (\ref{sixt}), the number of
integrations increases to 17. Besides that, averaging over
fluctuations of the refractive index introduces four additional
integrations. Finally, we have a $21$-fold integral. We have
performed most integrations analytically. The corresponding
procedure is similar to that described in Ref. \cite{ber}. The
rining, $5$-fold integral, has been calculated numerically. Fig. 1
shows the results for fixed values of the aperture ($r_0=4$) and
propagation distance ($z=10km$). We plot the dependence of the
dimensionless quantity $\langle R_w^2\rangle /R_b^2$ on the
turbulence strength.
\begin{figure}[t]
\centering
\includegraphics{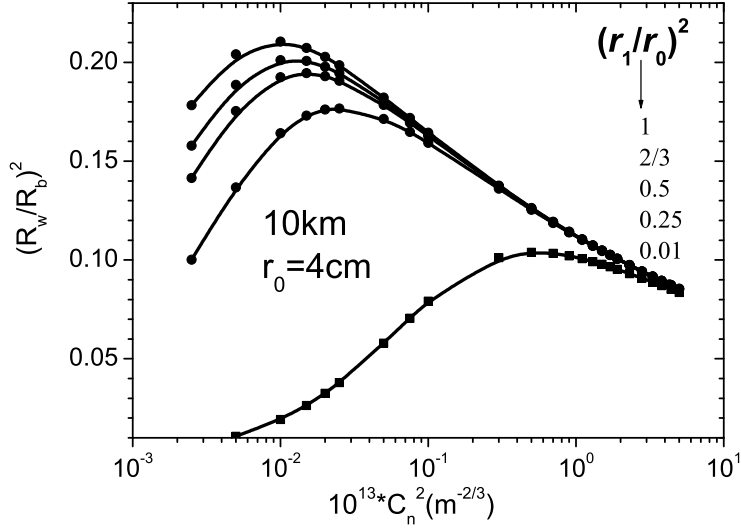}
\caption{Dimensionless mean square wandering radius  vs the
turbulence strength $C_n^2$ for different values of the partial
coherence determined by the ratio $r_1^2/r_0^2$. For all curves
$q_0=10^7m^{-1}, l^\prime_0=10^{-3}m$. The symbols show the results
of numerical calculations. For visual convenience, the solid lines
connect symbols using B-spline approximations.}
\end{figure}
(It is the ratio $\langle R_w^2\rangle /R_b^2$ rather than merely
$\langle R_w^2\rangle $ that is informative about the practical
significance of the wandering.) The beam radius $R_b$ is given by
the expression \cite{ber}
\begin{equation}\label{eigh}
R_b^2=\frac {r_0^2}2\bigg [1+\frac {4z^2}{q_0^2r_0^2r_1^2}+\frac {8z^3T}{r_0^2}\bigg ],
\end{equation}
where $T=0.558C_n^2l_0^{-1/3}$.

As we see in Fig. 1, there is still considerable beam wander even
for very strong turbulence, i.e. for $C_n^2=5*10^{-13}m^{-2/3}$.
(Usually, the value $C_n^2=10^{-14}m^{-2/3}$ is considered to be a
moderate turbulence level.) The four curves merge into a single
curve when $C_n^2\rightarrow \infty$. This means that the wandering
does not depend on the initial coherence in this case. So, the
universal behavior of the wandering effect corresponds to the
general concept that the atmosphere controls beam parameters for
long-distance propagation or for the strong turbulence regime. At
the same time, we see here the general tendency of the wander to
decay with the increasing turbulence strength supports the
reasonings of Fante \cite{fan}. He considered that when the
turbulence is strong, the beam no longer wanders significantly, but
rather breaks up into multiple beams.

In the opposite limiting case, $C_n^2\rightarrow 0$, the wander
distance $R_w$ also tends to zero  due to the obvious fact that the
wandering is entirely caused by turbulence. From a formal point of
view, there should be at least one maximum in the curve which
connects the regions of weak and strong turbulence. The
corresponding physical picture can be explained in terms of two
competitive tendencies occurring when $C_n^2$ increases: (i) in the
range of weak turbulence, where the beam radius is almost
independent of the turbulence, the probability to meet sufficiently
strong large-scale fluctuation of the refractive index, which
deflects the beam as a whole, increases linearly with $C_n^2$, (ii)
in the range of strong turbulence, there is considerable beam
widening due to photon scattering on fluctuations of the refractive
index ($R_b^2\sim C_n^2$) ; therefore the previous possibility has a
low probability. This explains the presence of the maxima in Fig. 1.

It is interesting to compare the results of the weak-turbulence
theory given by Eq. (\ref{twe}) with those of more general approach
based on the distribution function (\ref{fift}). The results are
shown in Fig. 2.
\begin{figure}[ht]
\centering
\includegraphics{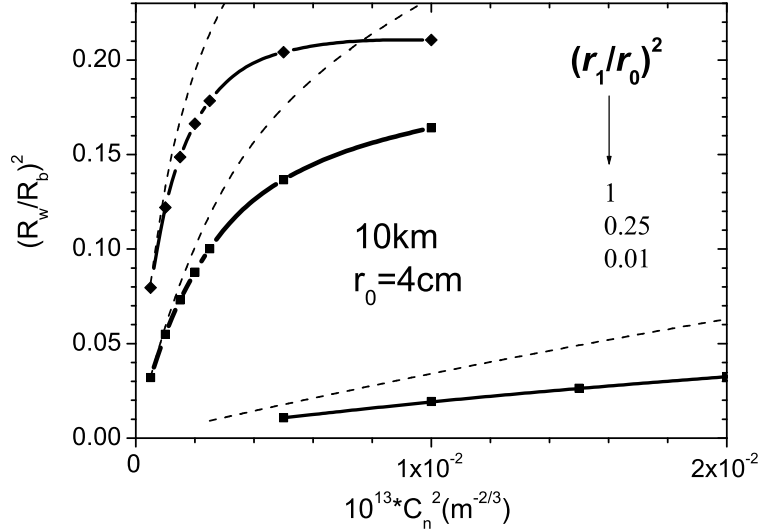}
\caption{Dependence of beam wandering on the turbulence strength for small values of $C_n^2$.
Dashed lines show the dependence given by Eq. (\ref{twe}), solid lines with symbols show
the results of more general theory.}
\end{figure}
As we see, both approaches give almost coinciding data for small
values of $C_n^2$. When $C_n^2$ increases, the results of the
weak-turbulence theory are overstated. A similar picture was
observed in Ref. \cite{bel} where the weak-turbulence theory was
tested by means of computer simulations.

Fig. 3 illustrates the dependence of beam wander on $C_n^2$ for a
shorter distance ($5km$) than in Fig. 1.
\begin{figure}[ht]
\centering
\includegraphics{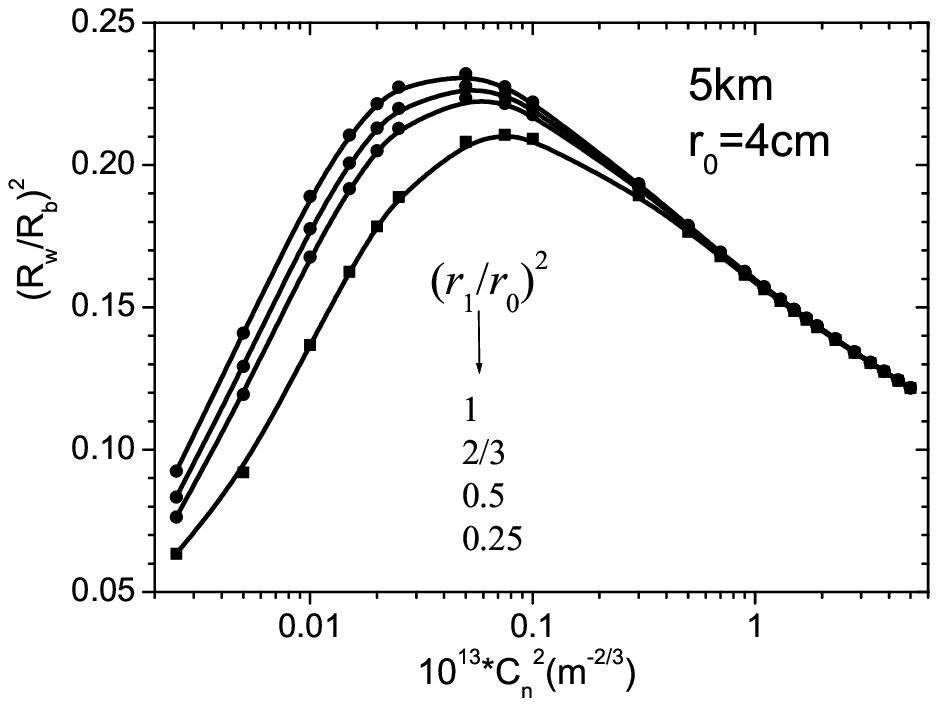}
\caption{The same as in Fig. 1 but for $z=5km$.}
\end{figure}
The plots in Figs. 1 and 3 are very similar with the only difference
that the maxima in Fig. 3 are displaced to the range of greater
values of the turbulence strength $C_n^2$. This difference is quite
evident. Namely, initially the overall effect of the turbulence
increases with the increase of both the value of $C_n^2$ and the
distance $z$. Therefore, the decrease of one of the factors can be
compensated by the increase of the other one, thus providing almost
the same effect of the turbulent atmosphere.
\begin{figure}[ht]
\centering
\includegraphics{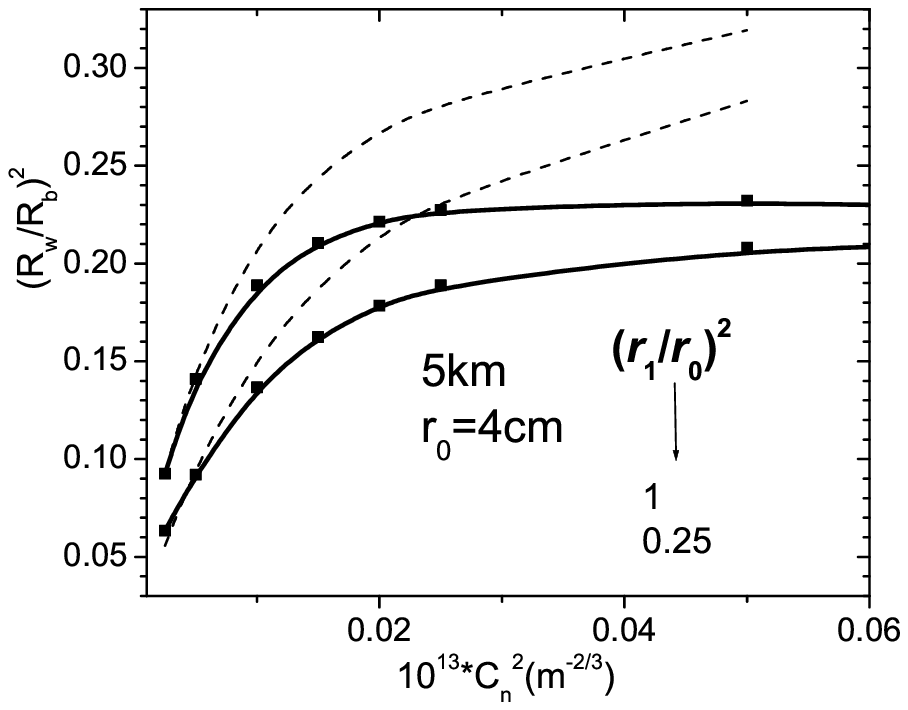}
\caption{The same as in Fig. 2 but for $z=5km$.}
\end{figure}
Fig. 4 illustrates how the two approaches correspond to one another
at small values of $C_n^2$. Again we see a good agreement of both
theories in this range of $C_n^2$.

Fig. 5 illustrates the dependence of the ratio $R_w^2/R_b^2$ on the
turbulence strength for small values of the aperture radius ($r_0=1
cm$).
\begin{figure}[ht]
\centering
\includegraphics{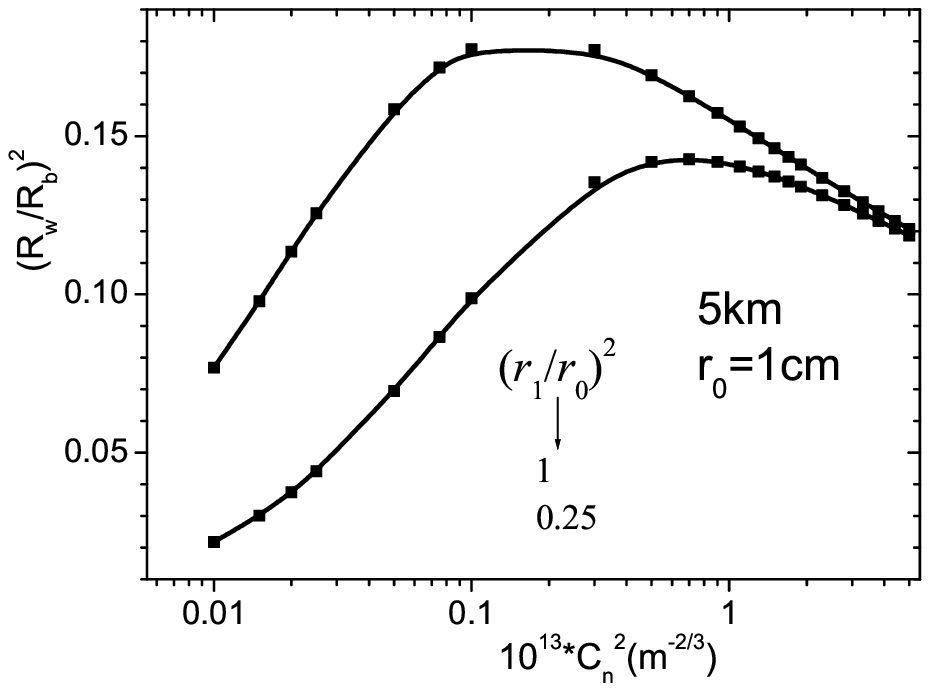}
\caption{The same as in Fig. 3 but for $r_0=1 cm$.}
\end{figure}
There is a significant decrease of the wandering effect in this
case. This is because a small value of $r_0$ (and automatically
$r_1$) results in considerable diffraction broadening of the beam
for such a long propagation path ($5 km$). That is why the influence
of turbulence on the beam parameters becomes competitive at greater
values of $C_n^2$ and, in correspondence with the latter, the maxima
of both curves are displaced to the right as compared with Fig. 3.
Also, the effect of partial coherence is more pronounced for smaller
initial radius of the beam. This can be seen by comparing Figs. 3
and 5.

The results presented in Figs. 1-5 require additional comments. Our
analysis proceeds from Eq. (\ref{two}) where the evolution of the
distribution function is based on the kinetic equation (\ref{fou}).
By definition this function is quadratic in the field amplitudes
 describing the intensity of the irradiance. Not all
momenta ${\bf q}_\perp $ and wave-vectors ${\bf k}_\perp $
contribute to the observable intensity. One can easily see that
initially in the absence of a phase screen, the characteristic
values of both $q_\perp$ and $k_\perp$ are given by $r_0^{-1}$. In
the presence of phase screen, the characteristic value of $q_\perp$
is of the order of $r_1^{-1}$ that determines the divergence of the
beam and its broadening (diffraction broadening). It follows from
geometric consideration that the diffraction broadening is of the
order of
 $z^2q_0^{-2}r_1^{-2}$. This is almost the same value as given by Eq. (\ref{eigh}). Therefore the
characteristic value of $k_\perp $ decreases with distance as
$(r_0^2/4+z^2 r_1^{-2}q_0^{-2})^{-1/2}$. Also, the momentum ${\bf
q}_\perp$ of the moving particle varies with distance due to
scattering on atmosphere inhomogeneities. The additional momentum
acquired in this way, $\Delta q _\perp$, can be estimated from its
mean square value as in the case of a Brownian particle moving in
${\bf q}_\perp $-space and being affected by a random force ${\bf
F}_\perp $ during the time $t=z/c $. Thus we have
\[ \langle\Delta q_\perp ^2\rangle\sim \langle F^2_\perp \rangle t\sim C_n^2z.\]  As a result,
the beam becomes more divergent and
additional broadening due to the turbulence, $\Delta R_b^2$, can be estimated as
\[\Delta R_b^2\approx \bigg\langle\bigg[\frac 1{q_0}\int _0^zdz^\prime \Delta q_\perp (z^\prime )\bigg
]^2\bigg\rangle \sim C_n^2z^3,\]. This again agrees with Eq.
(\ref{eigh}), where $\Delta R_b^2=2.23l_0^{-1/3} C_n^2z^3$.

When
\begin{equation}\label{a}
\Delta R_b^2>>\frac {r_0^2}{2}+\frac {2z^2}{q_0^2}r_1^{-2},
\end{equation}
one can say that the beam size is determined almost entirely by the
effects of the turbulence. In this case the characteristic values of
$k_\perp $ are of the order of $(\Delta R_b)^{-1}$ and decrease with
the increasing  turbulence  as $C_n^{-1}z^{-3/2}$. Also the
characteristic value of $q_\perp $ becomes of the order of $\Delta
q_\perp $ which is much greater than its initial value $r_0^{-1}$
(or $r_1^{-1}$). The last point can be seen directly from Eq.
(\ref{a}) when we represent the turbulence broadening as
\[ \Delta R_b^2\sim \frac {z^2}{q_0^2}\Delta q_\perp ^2.\]
 The condition $\Delta q_\perp >>r_0^{-1},r_1^{-1}$ means a considerable randomization of the radiation
field. The waves acquire  properties of Gaussian statistics that is
very important when calculating beam wander variance. In contrast to
calculations of the beam radius which is determined by correlations
of only two waves, the beam wandering effect is determined by
four-wave correlations (or by pair correlation function of the
intensity $\langle II\rangle $). The results presented in Figs. 1-5
were obtained explicitly assuming the dominating contribution to the
average
\[\langle I({\bf r}) I({\bf r}^\prime )\rangle\sim \sum_{\bf q,k}\sum_{\bf q^\prime ,k^
\prime }e^{-i({\bf kr}+{\bf k^\prime r^\prime})}\langle b^+_{{\bf
q}+{\bf k}/2} b_{{\bf q}-{\bf k}/2} b^+_{{\bf q}^\prime +{\bf
k}^\prime /2} b_{{\bf q}^\prime -{\bf k}^\prime /2}\rangle \] to be
from small regions of ${\bf k}$ and ${\bf k}^\prime $ as explained
above. (To simplify the notation, we omit the indices ($_\perp $) in
all variables.)  In this way strong correlations of pairs of waves $
b^+_{{\bf q}+{\bf k}/2}, b_{{\bf q}-{\bf k}/2}$ and $ b^+_{{\bf
q}^\prime +{\bf k}^\prime /2},  b_{{\bf q}^\prime -{\bf k}^\prime
/2}$ were taken into account. At the same time it is evident that
there is another region of wave vectors, i.e.
\[|{\bf q}+{\bf k}/2-{\bf q}^\prime +{\bf k}^\prime /2|, |{\bf q}-{\bf k}/2-{\bf q}^\prime -{\bf k}^\prime /2|
\sim \Delta R_b^{-1},\] where pair correlations of waves may be also
essential. The waves from different pairs, shown above, may
correlate in this region. Conventionally, we will refer to this type
of correlations as cross-correlation. In the case of strong
turbulence, the contribution of cross-correlations is not small,
thus providing saturation of fluctuations at high level. (See, for
example, \cite{ber}.) The two regions of wave vectors are well
separated from one another and possible overlapping in the course of
summing over wave vectors is not important in the case of strong
turbulence \cite{dash}. When the turbulence effect becomes weaker,
these regions approach each other, and in the limit of small
turbulence they unite into a single region. In this case the beam
wander is determined by the asymptotically exact solution
(\ref{twe}).

For strong turbulence, the contribution of cross correlations to the
beam wandering, $\langle R_w^2\rangle _{cross}$, can be obtained as
done in previous calculations. It is given by
  \begin{equation}\label{b}
\langle R_w^2\rangle _{cross}=\frac 83\frac {r_1^2}{r_0^2}\frac
{z^2}{q_0^2\Delta R_b^2}.
\end{equation}
Let us compare the value $\langle R_w^2\rangle_{cross}$ to $\langle
R_w^2\rangle $, shown in Figs. 1-5. First of all, consider those
$C_n^2$ which correspond to the maxima in the curves plotted in
Figs. 1,3,5.  For the case $r_0=r_1$ we see that $\Delta
R_b^2>>\frac {r_0^2}2, \frac {2z^2}{q_0^2r_0^2}$ in all cases. This
means that these maxima are in the range of strong turbulence, and
Eq. (\ref{b}) is applicable here.  The values obtained from Eq.
(\ref{b}) consist of only $7\%, 5\%, 0.4\%$ of the corresponding
data in Figs. 1,3,5, respectively. Moreover, if one moves towards
greater values of $C_n^2$, the contribution of cross-correlations
will become of smaller because of increasing of $\Delta R_b^2$. A
similar situation occurs when $r_1$ becomes less than $r_0$.

On the other hand, our solutions with cross-correlations neglected
almost coincide with those given by weak-turbulence theory when
$C_n^2\rightarrow 0$. (See Figs. 2 and 3.)  This assures us that
Figs. 1,3,5 represent reasonable solutions for the specific set of
parameters used there (and close to those) for any values of the
turbulence strength, $C_n^2$.

\section{Conclusion}
We have applied the method of photon distribution function
\cite{ber} to describe beam wander in turbulent atmosphere. In the
limit of weak turbulence and in the absence of artificial random
phase modulation, it becomes possible to obtain an analytical
expression for the wandering radius, which coincides with the one
known in the literature. Also, by bringing together analytical and
numerical calculations, we have succeeded in obtaining the wandering
radius in the range of strong turbulence. The general conclusion of
the actual studies is that the variation of the initial spatial and
temporal coherence provides significant positive (from the viewpoint
of practical applications of laser beams) influence on the character
of the intensity fluctuations. Namely, the relative value of the
wandering radius can be considerably reduced. Moreover this
reduction takes place just in the range of the most pronounced
wandering effect. (See Figs. 1,3,5.) At the same time, the effect of
partial coherence vanishes for very strong turbulence. This is in
contrast to the behavior of the scintillation index, which in this
case can be significantly suppressed by decreasing the initial
coherence of the light. (See, for example, Refs. \cite{ban} and
\cite{ber}.) But this suppression  is not very important because of
small wandering effect in this case.

\section{Acknowledgment}
We thank B.M. Chernobrod, G.D. Doolen, and P.W. Milonni for
discussions. This work was carried out under the auspices of the
National Nuclear Security Administration of the U.S. Department of
Energy at Los Alamos National Laboratory under Contract No.
DE-AC52-06NA25396.

\newpage \parindent 0 cm \parskip=5mm





\end{document}